\documentclass[11pt,twocolumn,secnumarabic,aps,amsmath,amssymb,amsfonts,superscriptaddress,tightenlines,nobibnotes,prd,nofootinbib]{revtex4}

\usepackage{bm}
\usepackage[utf8]{inputenc}
\usepackage{amsmath}
\usepackage{amssymb}
\usepackage{bbold}
\usepackage{pgfplots}
\usepackage{wasysym}
\usepackage{subeqnarray}
\usepackage{graphicx}
\usepackage{color}
\usepackage{siunitx}

\usepackage{epsfig}
\usepackage[hidelinks,colorlinks=true,citecolor=blue,linkcolor=blue]{hyperref}

\numberwithin{equation}{section}
\newcommand{\be}{\begin{equation}}
\newcommand{\ee}{\end{equation}}
\renewcommand\[{\begin{equation}}
\renewcommand\]{\end{equation}}
\newcommand{\obs}{{\rm o}}
\newcommand{\src}{{\rm s}}
\newcommand{\dd}{{\rm d}}
\newcommand{\myZ}{{\cal W}}
\newcommand{\myE}{{\cal E}}
\newcommand{\myB}{{\cal H}}

\begin{document}

\title{Direction and redshift drifts for general observers and their applications in cosmology}

\author{Oton H. Marcori}
\email[]{otonhm@hotmail.com}
\affiliation{Departamento de Física, Universidade Estadual de Londrina, Rod. Celso Garcia Cid, Km 380, 86057-970, Londrina, Paraná, Brazil.}
\author{Cyril Pitrou}
\email[]{pitrou@iap.fr}
\affiliation{Sorbonne Universit\'e, CNRS, UMR 7095, Institut d’Astrophysique de Paris, 98 bis bd Arago, 75014 Paris, France.}
\author{Jean-Philippe Uzan}
\email[]{uzan@iap.fr}
\affiliation{Sorbonne Universit\'e, CNRS, UMR 7095, Institut d’Astrophysique de Paris, 98 bis bd Arago, 75014 Paris, France.}
\author{Thiago S. Pereira}
\email[]{tspereira@uel.br}
\affiliation{Departamento de Física, Universidade Estadual de Londrina, Rod. Celso Garcia Cid, Km 380, 86057-970, Londrina, Paraná, Brazil.}
\date{\today}
\begin{abstract}
High precision astrometry now enables to measure the time drift of astrophysical observables in real time, hence providing new ways to  probe different cosmological models. This article presents a general derivation of the redshift and direction drifts for general observers. It is then applied to the standard cosmological framework of a Friedmann-Lema\^{\i}tre spacetime including all effects at first order in the cosmological perturbations, as well as in the class of spatially anisotropic universe models of the Bianchi~I family. It shows that for a general observer, the direction drift splits into a parallax and an aberration drifts and order of magnitude estimates of these two components are provided. The multipolar decomposition of the redshift and aberration drifts is also derived and shows that the observer's peculiar velocity contributes only as a dipole whereas the anisotropic shear contributes as a quadrupole.
\end{abstract}
\maketitle


\section{Introduction}

Over the past decades, the developments of observational cosmology, among which cosmic microwave background (CMB), baryonic acoustic oscillations (BAO), type Ia supernovae and cosmic shear, have led to a robust model of our universe in the framework of Friedmann-Lema\^{\i}tre spacetimes, the parameters of which are well-measured to define the \emph{concordance model} of cosmology. The need for a dark sector has triggered the necessity to develop tests of the hypothesis on which this model lies (see Refs.~\cite{Uzan:2009mx,Uzan:2016wji} for their description and existing tests). Any data which is not strictly located on our past light cone brings sharp constraints, in particular on the Copernican principle~\cite{Dunsby:2010ts}.

The first evaluation of the time drift of the cosmic redshift dates back to the 60s~\cite{1962ApJ...136..319S,McVittie}. Since it offers a direct measurement of the local cosmic expansion rate~\cite{Loeb:1998bu}, it evolved to the idea of ``real time'' cosmology~\cite{Quercellini:2010zr}, based on the time drifts of both the cosmic redshift and the direction as new cosmological observables. For example, the possibility of measuring the redshift drift was thought both as a way of measuring the instantaneous cosmic expansion rate as a function of redshift, $H(z)$,~\cite{2008MNRAS.386.1192L} and hence better constrain dark energy models~\cite{2009MNRAS.397.1739D,1988ApJ...331..648R,doi:10.1143/PTP.79.777} in the framework of Friedmann-Lema\^{\i}tre cosmologies. On a more fundamental level,  the redshift drift was shown to offer a way to test of the Copernican principle in~\cite{Uzan:2008qp,Dunsby:2010ts,2011PhRvD..83d3527Y}, while lensing-type effects of a local void were investigated in~\cite{Cusin:2016kqx}. Direction drift effects were also investigated in the test of the Copernican principle in Refs.~\cite{2010PhRvD..81d3522Q}, while in Refs.~\cite{2009PhRvD..80f3527Q,2009PhRvD..80l3515F} it was used to constrain anisotropic (Bianchi~I) cosmological models. 
More recently, the aberration drift was used as a redshift-independent tool to extract the proper acceleration of the earth with respect to CMB~\cite{2018arXiv180204495B}. From a more formal perspective,  several investigations of optical drift effects in general relativity provide a covariant derivation of cosmic parallax for a pair of sources in general spacetimes~\cite{low1993celestial,perlick1990redshift,hasse1988geometrical,Rasanen:2013swa,Korzynski:2017nas}.

From an observational perspective, the developments of precision astrometry have allowed the Gaia space mission to measure the parallax of astrophysical objects with unprecedented precision~\cite{2006MNRAS.367..290J,2018arXiv180409376L}. Forthcoming experiments carrying state-of-the-art spectrography on the E-ELT, aim to reach the sensitivity to measure the redshift drift by monitoring Ly-$\alpha$ absorption lines of distant quasars in a time span of a decade~\cite{pasquini2005codex} and it has been demonstrated that the use of many spectral lines and quasars is an important measure to reduced the variance in $\dot{z}$ induced by linear cosmological perturbations~\cite{Uzan:2007tc}. 

This article revisits both the redshift and direction drifts in several cosmological frameworks. It starts by a simple analysis in Minkowski spactime in Section~\ref{Sec2} in order to emphasize the need for a two-worldlines analysis, hence non-local in time. A heuristic argument allows one to grasp the physical meaning and typical amplitude of all the different contributions. It takes into account a general accelerated observer and source. Then, Section \ref{Sec3} focuses on the computation of the redshift and direction drifts for general observers in FL spacetime. A multipolar decomposition of this result is presented and  compared to the literature. The order of magnitude of the aberration and parallax parts of the direction drift are then estimated. In Section \ref{Sec3b}, the same drift effects are computed but now for a linearly perturbed FL spacetime, including scalar, vector and tensor perturbations. This fully characterizes the imprints of the large scale structure. The analysis is finally extended in Section~\ref{sec:Bianchi-I} to anisotropic spacetimes to provide the expressions of these drifts for a Bianchi I universe. As such it offers a new way to test the isotropy of the cosmic expansion around us, complementary to methods based on either cosmic shear~\cite{Fleury:2014rea,Pitrou:2015iya,Pereira:2007yy} or supernovae~\cite{Appleby:2012as,Bengaly:2015dza} observations. Section \ref{Sec5} summarizes our results and their implications.

\section{General approach} \label{Sec2}

\subsection{Heuristic argument}

In order to get some intuition on the aberration drift, let us start the computation with a heuristic approach.  In 1728, Bradley~\cite{bradley} obtained an expression for stellar aberration using a corpuscular description of light with Newton optics theory. It is today well-understood in the framework of special relativity that aberration is related to the Lorentz boost associated with a change of referential (see e.g. Ref.~\cite{dubook} for details). Consider an inertial frame $S$ and a second inertial frame $S'$ moving with speed $V$ with respect to $S$ along the $X$ axis. For a light source in the XOY plane of $S$ whose radius vector makes an angle $\alpha$ with the OX axis, the wave-vector of the light emitted by the star is $k^\mu=k^0(1,\cos\alpha,\sin\alpha,0)$. It is $k^{\prime\mu}=k^{\prime 0}(1,\cos\alpha',\sin\alpha',0)$ in $S'$. It is a standard exercise to show that the angles are related by
\be
 \tan\alpha'=\frac{\sqrt{1-V^2}}{1-V/\cos\alpha}\tan\alpha\,.
\ee
 or equivalently
 \be
 \cos\alpha' = \frac{\cos\alpha-V}{1-V\cos\alpha}\,,
 \ee
 where the speed of light has be set to unity by a proper choice of units. Bradley considered $S$ to be the quasi-inertial frame of the Solar system and $S'$ attached to the Earth. Over a $\Delta t=6$~months period, $\Delta V=2V$ so that the aberration is given by
 \[
 |\cos(\alpha'+\Delta\alpha')-\cos\alpha'| \sim \sin\alpha' \Delta\alpha'\sim 2V\sin^2\alpha'. \nonumber
 \]
 The aberration is just an effect of perspective due to a change of (Lorentz) frame, and the question of whether or not to take into account the source velocity does not arise, which was however a difficult question in corpuscular optics. It follows that the drift is of order
 \[
 \frac{\Delta\alpha'}{\Delta t} \sim \frac{\Delta V}{\Delta t}\sin\alpha'\,.
 \]
 
There is however a second effect to take into account.  It is a simple geometric effect due to the fact that in $\Delta t$ the star and the observer have moved by a typical relative distance $\vert {\bf v}_* -  {\bf v}_\obs\vert\Delta t$ so that it is expected that
\[
\Delta\alpha'\sim\frac{\vert {\bf v}_* -  {\bf v}_\obs\vert\Delta t}{\vert {\bf r}_* -  {\bf r}_\obs\vert}.
\]
In conclusion, we expect the typical total drift to behave as
\[
 \frac{\Delta\alpha'}{\Delta t} = \frac{\vert {\bf v}_* -  {\bf v}_\obs\vert}{\vert {\bf r}_* -  {\bf r}_\obs\vert} + \frac{\Delta V}{\Delta t} \vert \sin\alpha\vert
\]
We shall refer to the total drift as \emph{direction drift}, the term inversely proportional to the source-observer distance is the \emph{parallax drift} and the term proportional to the observer's acceleration is the \emph{aberration drift}. Indeed, the arguments above are just a simple sketch of the different contributions. The angles have to be replaced by direction on the celestial sphere (i.e., unit 2-vectors) and we need to properly define the angles, motions, wave-vectors, etc., as described by non-inertial observers and sources. However, it emphasizes a key issue on which our formalism is built: the direction drift, and similarly the redshift drift, involves two sets of null-geodesic compared before and after a given lapse of observer proper time. It also shows that the direction drift cannot be reduced to a time derivative (coordinate or proper) of the aberration.

\subsection{Minkowski spacetime}

Let us consider a Minkowski spacetime with metric $\dd s^2=-\dd t^2+\delta_{ij}\dd x^i\dd x^j$ in Cartesian coordinates. Thanks to the three spatial translations Killing vectors, ${\xi^\mu_{(i)}=\delta^\mu_i}$, the vector $k^\mu$ tangent to a null geodesic $x^\mu(\lambda)$ satisfies
\begin{equation}\label{00cond}
k_{\mu} \xi^{\mu}_{(i)} = k_{i}=\text{cte}
\end{equation}
\emph{along the geodesic}.
Likewise, the existence of a timelike Killing vector ensures that the frequency measured by the associated observer is constant along the photon's trajectory. 

Let us first consider a static observer with 4-velocity $u^\mu \equiv \delta_0^\mu$. We decompose $k^\mu$ into a frequency $\omega$ and a direction $n^\mu$ as
\begin{equation}
 k^{\mu}=\omega(u^{\mu}-n^{\mu}), \qquad n^\mu u_\mu = 0.
 \end{equation}
The condition~(\ref{00cond}) reduces to
\begin{equation} \label{eq:geodfl1M4}
 \frac{\dd x^{i}}{\dd t}=- n_\obs^i\,,
\end{equation}
and we have fixed the constant on the r.h.s. of Eq.~\eqref{eq:geodfl1M4} to be the value of the vector $n^i$ at the observer's position. This equation is easily integrated to give
\begin{equation} \label{eq:geodfl2M4}
x^{i}_{s}-x^{i}_\obs=-(t_s - t_\obs) n_{\obs}^i\,,
\end{equation}
where the subscripts ``s" and ``o" stand for ``source" and ``observer" respectively. Let us now consider a second geodesic in which the photon was emitted at $t_s+\delta t_s$ and received by the observer at $t_\obs+\delta t_\obs$. We also assume that the source is static. Writing Eq.~\eqref{eq:geodfl1M4} for this second geodesic gives
\begin{equation} \label{eq:geodM2}
x^{i}_{s}-x^{i}_\obs=-\left(t_s - t_\obs +\delta t_s-\delta t_\obs \right) \left( n_{\obs}^i + \delta_{12}n_{\obs}^i \right),
\end{equation}
where we have introduced the general definition 
\begin{equation}
\delta_{12} {\cal O} = {\cal O}(t_\obs+\delta\tau) - {\cal O}(t_\obs),
\end{equation}
for any observable ${\cal O}$ compared at two times separated by a lapse $\delta\tau$ of the proper time of the observer. Note that it is a non-local variation which involves two geodesics. For a static observer, we also have $\delta\tau=\delta t$. Contraction of Eqs.~\eqref{eq:geodfl1M4} and~\eqref{eq:geodM2} with $n_{\obs}^i$ leads to $\delta t_s=\delta t_\obs$, as expected\footnote{Note that we have also used $n_{i\obs}\delta_{12}n^i_\obs=0$, which is true for small variations.}. If we now compare Eqs.~\eqref{eq:geodfl2M4} and~\eqref{eq:geodM2} to extract the direction drift, we find for a static source and a static observer in Minkowski spacetime
\begin{equation}
\delta_{12}n_\obs^i=0,
\end{equation}
i.e., there is no direction drift.

Let us now consider a general observer with 4-velocity $\tilde{u}^{\mu}$. We assume that its peculiar velocity $v^\mu$ is small compared to the speed of light. Hence, we shall work at linear order in the spatial velocity and write
\begin{equation}
 \tilde{u}^{\mu}=u^{\mu}+v^{\mu},
\end{equation}
with $v^{\mu}u_\mu=0$. From now on tilde quantities are associated to the general observer. The wave-vector is either decomposed as $k^{\mu}=\omega(u^{\mu}-n^{\mu})$ for the static observer, or as $k^{\mu}=\tilde{\omega}(\tilde{u}^{\mu}-\tilde{n}^{\mu})$ for the general observer. It follows that
\begin{subeqnarray}\label{LabGlobal}
k^{i}&=&\tilde{\omega}\left(v^{i}-\tilde{n}^{i}\right), \slabel{accobs1-M4} \slabel{Sub1}\\
\tilde{\omega}&=&\omega\left(1+v^{i} n_{i}\right),\slabel{accobs2-M4} \\
\tilde{n}^{i}&=&n^{i}+\bot^{i}_{j}v^{j}, \slabel{accobs3-M4}
\end{subeqnarray}
where
\begin{equation}\label{Defbot}
 \bot_{j}^{i} \equiv \delta_{j}^{i}-n^{i}n_{j}
 \end{equation}
is the perpendicular projector with respect to the direction of observation of the static observer, $n^{i}$. The proper time of the general observer is related to the coordinate time by $\dd \tau^2=(1-v^2)\dd\tilde{t}^2$, hence $\dd \tau = \dd \tilde{t}$ at linear order in $v$. Since Eq.~\eqref{00cond} is observer independent, the geodesic equation leads to 
\begin{equation}
 \omega \frac{\dd x^{i}}{\dd t}= \tilde{\omega}_\obs\left(v_\obs^{i}-\tilde{n}_\obs^{i}\right).
\end{equation}
Using Eq.~\eqref{accobs2-M4}, it takes the form
\begin{equation} \label{geodeqacc1-M4}
 \frac{\dd x^{i}}{\dd t}=\bot_{j}^{i}v_\obs^{j}-\tilde{n}_\obs^{i}\,.
\end{equation}
In this expression, we have used that the perpendicular projector multiplies the spatial velocity, and therefore the relation $n^i = \tilde n^i$, valid at lowest order in $v$, can be used in Eq.~(\ref{Defbot}). Once integrated, it leads to
\begin{equation} \label{geodeqacc3-M4}
  x^{i}_{s}-x^{i}_{\obs}=(t_s - t_\obs)\left( \bot_{j}^{i}v_\obs^{j}-\tilde{n}_\obs^{i}  \right)
\end{equation}
since $n^i$ and thus $\bot_{j}^{i}=\delta_{j}^{i}-n^{i}n_{j}$  are constants along the null geodesic. This is the generalization of Eq.~(\ref{eq:geodfl2M4}) for a general observer.

\begin{figure}[h!]
\centering
\includegraphics[width=.6\columnwidth]{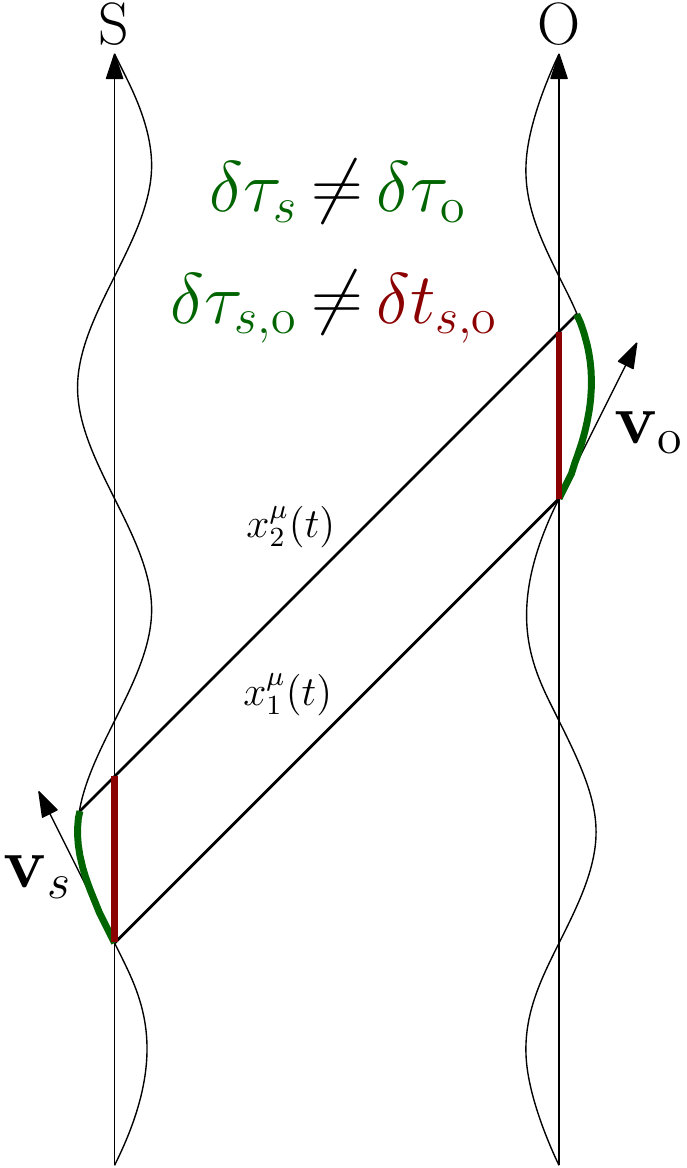}
\caption{General definitions for the  two null geodesics between a general observer and source. The differences between the times contribute to the parallax part of the drift. Note that at first order in velocities $\delta\tau_{s,\obs}=\delta \tilde{t}_{s,\obs}$.} \label{fig1}
\end{figure}

Let us now consider a second light ray emitted at $t_s+\delta\tilde{t}_s$ and received at $t_\obs+\delta\tilde{t}_\obs$. We also allow the source to have a general velocity. We now have $\delta\tilde{t}_\obs\not=\delta\tilde{t}_{s}$, since both the observer and source are moving. The end points of the second null-geodesic are now given by  $x^{i}_{s,\obs}+v^{i}_{s,\obs}\delta\tilde{t}_{s,\obs}$, and the velocity of the observer is then $v_{\obs}^{i}+\dot v_{\obs}^{i}\delta\tilde{t}_\obs$. Figure\ref{fig1} illustrates the scheme of this calculation. The integration of the second geodesic gives
\begin{equation}
\begin{split} \label{geodeqacc2-M4}
&x^{i}_{s}-x^{i}_\obs+v^{i}_{s}\delta\tilde{t}_{s}-v^{i}_{\obs}\delta\tilde{t}_{\obs}=\\
&\qquad \left(t_s - t_\obs +\delta\tilde{t}_{s}-\delta\tilde{t}_{\obs}\right)\times \\
&\qquad \quad \left(\bot_{j}^{i}v_{\obs}^{j}-\tilde{n}^{i}_{\obs}+\bot_{j}^{i}\dot v_{\obs}^{j}\delta\tilde{t}_{\obs}-\delta_{12}\tilde{n}^{i}_{\obs}\right).
\end{split}
\end{equation}
The relation between $\delta\tilde{t}_{s}$ and $\delta\tilde{t}_\obs$ is obtained by subtracting the contractions of Eqs.~\eqref{geodeqacc3-M4} and \eqref{geodeqacc2-M4} with $n^{i}_{\obs}$ to get
\begin{equation}\label{deltat-so}
  \frac{\delta\tilde{t}_{s}}{\delta\tilde{t}_{\obs}}=1-\left(v^{i}_{s}-v^{i}_{\obs}\right)n_{\obs i}.
\end{equation}
Once inserted in  Eqs.~\eqref{geodeqacc3-M4} and~\eqref{geodeqacc2-M4}, we get the direction drift
\begin{equation} \label{abdriftacc-M4}
\frac{\delta_{12}\tilde{n}^{i}_{\obs}}{\delta t_{\obs}}=\frac{\bot_{j}^{i}\left(v^{j}_{s}-v^{j}_{\obs}\right)}{D_{s\obs}}+\bot_{j}^{i}\dot{v}_{\obs}^{j},
\end{equation}
where $D_{s\obs}\equiv t_\obs - t_s$ is the (positive) radial distance between the observer and the source. Note that the perpendicular projector multiplies spatial velocities, and as we work at linear order in velocities, it is equivalent to consider the projector built either from $n^i$ or $\tilde n^i$.

The expression~(\ref{abdriftacc-M4}) matches the heuristic argument. The first term (the parallax drift) is related to the change of apparent position of the source due to the motion of the source and the observer. This is a simple perspective effect that is proportional to the inverse of the distance. The second term (the aberration drift) is the mean peculiar acceleration of the observer perpendicular to the line of sight on the time scale of the observation. 

\subsection{Summary}

This first insight shows that it is indeed not correct to start from the Bradley formula for the aberration and simply take its time derivative. It also defines clearly the general strategy of the computation: (1) consider 2 null geodesics connecting the observer and the source, (2) relate the lapses of time between the two geodesics at the source and at the observer. This provides a well-defined and physically well-under control derivation of both the direction and redshift drifts that can be extended to any spacetime. It also emphasizes that we have to pay attention to the fact that $\omega$ may not be constant along the geodesics and to define properly the direction of observation. To that purpose, given a metric $g_{\mu\nu}$, we shall introduce a tetrad basis $\epsilon^\mu_{(A)}$ defined by
\begin{equation}
  g_{\mu\nu}\epsilon^\mu_{(A)}\epsilon^\nu_{(B)} = \eta_{AB}.
\end{equation}
If we choose $\epsilon_{(0)}^\mu$ to be the 4-velocity $u^\mu$ of an inertial observer, any vector field $V^\mu$ can then be decomposed as
\begin{equation}
 V^\mu = (-V^\alpha u_\alpha) u^\mu + V^{(i)}\epsilon^\mu_{(i)}.
\end{equation}

\section{Friedmann-Lema\^{\i}tre spacetime and its perturbations} \label{Sec3}

\subsection{General observers}

The construction from the previous section is easily generalized to a cosmological spacetime. Let us assume a cosmology described 
by a spatially flat Friedmann-Lema\^{\i}tre (FL) solution with metric
\begin{equation}
\dd s^2 = -\dd t^2 + a^2(t)\delta_{ij}\dd x^i \dd x^j
\end{equation}
where $t$ is the cosmic time, i.e. the proper time of comoving observers with 4-velocity $u^\mu\equiv \delta_0^\mu$ and $a(t)$ is the scale factor. We also introduce the conformal time defined by $\dd\eta=a^{-1}\dd t$ and define the Hubble function by $H\equiv \dd\ln a/\dd t$. We use the tetrad field defined by
\begin{equation}\label{tetrad00}
\bm{\epsilon}_{(0)}=\partial_{t}, \qquad 
\bm{\epsilon}_{(i)}=a^{-1}\partial_{i}\,.
\end{equation}
For a photon with wave-vector ${k^{\mu}=\omega\left(u^{\mu}-n^{\mu}\right)}$, one easily derives from the formalism of the previous sections that the redshift drift for comoving observers and sources is
\begin{equation}
\frac{\delta_{12}  z}{\delta t_{\obs}}=(1+{z})H_{\obs}-H_s,
\end{equation}
whereas the direction drift vanishes,
\begin{equation} \label{abdriftinobs}
\delta_{12} {n}^{(i)}_\obs=0
\end{equation}
when $n^i$ is expressed in its tetrad components $n^i=n^{(j)}\epsilon^i_{(j)}$. The case of comoving observer and sources is indeed  analogous to the Minkowski case for static observer and sources.

As in the previous section, we now consider a general observer with 4-velocity
\begin{equation}
  \tilde{u}^{\mu}=u^{\mu}+v^{\mu},
\end{equation} 
where $v^{\mu}$ is spatial ($v^\mu u_\mu=0$). Again,  the tilde denotes quantities associated with a general observer. We assume the observer is non-relativistic and work at first order in $v^\mu$.

The wave-vector is decomposed equivalently either as $k^{\mu}=\omega(u^{\mu}-n^{\mu})$ or as $k^{\mu}=\tilde{\omega}(\tilde{u}^{\mu}-\tilde{n}^{\mu})$. Again, it follows that
\begin{subeqnarray}
k^{i}&=&\tilde{\omega}\left(v^{i}-\tilde{n}^{i}\right), \slabel{accobs1} \\
\tilde{\omega}&=&\omega\left(1+v^{i}n_{i}\right), \slabel{accobs2} \\
\tilde{n}^{i}&=&n^{i}+\bot^{i}_{j}v^{j}, \slabel{accobs3}
\end{subeqnarray}
where $\bot_{j}^{i}=\delta_{j}^{i}-n^{i}n_{j}$ is the perpendicular projector with respect to $n^{i}$. Since the FL spacetime also enjoys 3 spatial Killing vectors,
\begin{equation}\label{xxxki}
a^2k^{i} = k_i = {\rm cte}
\end{equation}
along a null geodesic. However, FL having nly a timelike conformal Killing vector, we have ${a\omega=\text{cte}}$. After decomposing $n^i$, $\tilde n^i$ and the velocities on the tetrad~(\ref{tetrad00}) as 
\begin{equation}
 n^i = n^{(j)} \epsilon^i_{(j)}, \quad
 \tilde n^i = \tilde n^{(j)} \epsilon^i_{(j)}, \quad
  v^i = v^{(j)} \epsilon^i_{(j)}, 
\end{equation}
the condition~(\ref{xxxki}) leads to
\begin{equation}
a\omega \frac{\dd x^{i}}{\dd\eta}=a_{\obs}\tilde{\omega}_{\obs}\left(v_{\obs}^{(i)}-\tilde{n}_{\obs}^{(i)}\right).
\end{equation}
With the use of Eq.~\eqref{accobs2}, and the fact that ${a\omega=a_\obs\omega_\obs}$ for comoving observers, it reduces to
\begin{equation} \label{geodeqacc1}
\frac{\dd x^{i}}{\dd\eta}=\bot_{j}^{i}v_{\obs}^{(j)}-\tilde{n}_{\obs}^{(i)}.
\end{equation}
It can be integrated easily since $n^{(i)}$ is constant along the null geodesic, and so is $\bot_{j}^{i}$, once expressed as 
${\bot_{j}^{i}=\delta^i_j -n^{(i)}n_{(j)}}$. We get
\begin{equation} \label{geodeqacc3}
x^{i}_{s}-x^{i}_{\obs}=\left( \eta_{s}-\eta_{\obs} \right) \left(  \bot_{j}^{i}v_{\obs}^{(j)}-\tilde{n}_{\obs}^{(i)}  \right).
\end{equation}
Indeed, after shifting to conformal time, this is similar to the derivation in Minkowski spacetime.

We now shift to the second geodesic describing the observation at a conformal time $\eta_{\obs}+\delta\tilde{\eta}_{\obs}$ where $\delta\tilde{\eta}_{\obs}$ is related to the proper time $\tau$ of the observer by $\delta\tau=\delta\tilde{t}$ as long as we work at first order in velocities. The photon was emitted by the source at a conformal time $\eta_{s}+\delta\tilde{\eta}_{s}$.  As explained in the previous section, we need once more to find the relationship between $\delta\tilde{\eta}_s$ and $\delta\tilde{\eta}_\obs$. Compared to the first geodesic, the positions of the source and the observer are now given by  $x^{i}_{s,\obs}+v^{(i)}_{s,\obs}\delta\tilde{\eta}_{s,\obs}$; the velocity of the observer is $v_{\obs}^{(i)}+v_{\obs}^{\prime(i)}\delta\tilde{\eta}_{\obs}$ (where a prime refers to a derivative with respect to the conformal time). As long as we work in first order in velocities, we need only the background value of  $\delta_{12} n^{(i)}_{\obs}=0$ so that $\bot_{j}^{i}$  remains unchanged. Plugging these modifications in Eq.~(\ref{geodeqacc1}) we get the equation of the second geodesic
\begin{align} \label{geodeqacc2} 
&x^{i}_{s}-x^{i}_{\obs}+v^{(i)}_{s}\delta\tilde{\eta}_{s}-v^{(i)}_{\obs}\delta\tilde{\eta}_{\obs}
=\\
&\quad\quad\left(\eta_{s}-\eta_{\obs}+\delta\tilde{\eta}_{s}-\delta\tilde{\eta}_{\obs}\right)\times \nonumber \\
&\quad\times\left(\bot_{j}^{i}v_{\obs}^{(j)}-\tilde{n}^{(i)}_{\obs}+\bot_{j}^{i}v_{\obs}^{\prime(j)}\delta\tilde{\eta}_{\obs}-\delta_{12}\tilde{n}^{(i)}_{\obs}\right). \nonumber
\end{align}
By contracting this latter equation with $n^{(i)}_{\obs}$, one gets the relation between $\delta\tilde{\eta}_{s}$ and $\delta\tilde{\eta}_{\obs}$ as
\begin{equation}\label{deltaeta-so}
 \frac{\delta\tilde{\eta}_{s}}{\delta\tilde{\eta}_{\obs}}=1-\left(v^{(i)}_{s}-v^{(i)}_{\obs}\right)n_{(i)}.
\end{equation}
This can also be obtained by making use of Eq.~\eqref{tzrelation} derived in Appendix~A.

To finish, we just need to combine Eqs.~\eqref{geodeqacc3}, \eqref{geodeqacc2} and \eqref{deltaeta-so} to get the total direction drift, i.e., the change of direction 
in units of proper time of the observer, which is
\begin{equation} \label{abdriftgenobs}
\frac{\delta_{12}\tilde{n}^{(i)}_{\obs}}{\delta t_{\obs}}=\frac{\bot_{j}^{i}\left(v^{(j)}_{s}-v^{(j)}_{\obs}\right)}{a_{\obs}\chi_{s\obs}}+\bot_{j}^{i}\dot{v}_{\obs}^{(j)},
\end{equation}
where $\chi_{s\obs}\equiv\eta_{\obs}-\eta_{s}$ is the comoving radial distance between the observer and the source. This expression is similar to Eq.~\eqref{abdriftacc-M4} derived in a Minkowski spacetime, and its implications will be discussed in~\S~\ref{disc-velocities}.

Let us now turn to the redshift drift. Starting from Eq.~\eqref{accobs2}, we can express the redshift as
\[
1+\tilde{z}=\frac{\tilde{\omega}_{s}}{\tilde{\omega}_{\obs}}=\frac{a_{\obs}}{a_{s}}\left[1+\left(v^{(i)}_{s}-v^{(i)}_{\obs}\right)n_{(i)}\right].
\]
It is then straightforward to write the redshift $1+\tilde{z}+\delta \tilde{z}$ for a second geodesic and subtract the previous relation, so as to find the observed redshift drift,
\begin{align}\label{zdot}
\frac{\delta_{12} \tilde{z}}{\delta t_{\obs}}=&(1+\tilde{z})\left(H_{\obs}-\dot{v}^{(i)}_{\obs}n_{(i)}\right)\nonumber\\
&-\left(H_{s}-\dot{v}^{(i)}_{s}n_{(i)}\right).
\end{align}

\subsection{Multipolar decomposition}\label{SecMultipoles1}

The angular dependence, i.e. in $n^{(i)}$, of the redshift and aberration drifts can be decomposed in multipoles either using symmetric trace-free tensors or spherical harmonics~\cite{Thorne1980}. For the redshift drift, it takes the form
\begin{align}
\frac{\delta_{12} \tilde{z}}{\delta t_{\obs}} &= \myZ + \myZ_i n^{(i)} + \myZ_{ij} n^{(i)} n^{(j)}+\dots\nonumber\\
& = \sum_{\ell, m} \myZ_{\ell m} Y^{\ell m}(n^{(i)}) 
\end{align}
where the $\myZ_{i_1\dots i_n}$ are symmetric trace-free tensors.  From Eq. (\ref{zdot}), we read that the monopole and the dipole of the redshift drift are
\begin{subeqnarray}
\myZ&=&(1+\tilde{z})H_{\obs}-H_s\\
\myZ_i&=&-(1+\tilde{z})\dot{v}^{(i)}_{\obs} +\dot{v}^{(i)}_{s}\,.
\end{subeqnarray}

The direction drift can be written in terms of a gradient and curl as
\begin{equation}
\frac{\delta_{12}\tilde{n}^{(i)}_{\obs}}{\delta t_{\obs}} = D^{(i)} \myE(n^{(l)}) +
\epsilon^{(i)(j)} D_{(j)} \myB(n^{(l)})
\end{equation}
where $D^{(i)}$ is the covariant derivative on the 2-sphere of unit radius and $\epsilon^{(i)(j)}$ is the Levi-Civita tensor on the unit 2-sphere. The angular dependence of $\myE(n^{(i)})$ and $\myB(n^{(i)})$ is in turn decomposed in symmetric trace-free tensors or in spherical harmonics. For $\myE$ (and similarly for $\myB$) this decomposition is
\begin{align}
\myE(n^{(l)}) &= \myE_i n^{(i)} +\myE_{ij} n^{(i)} n^{(j)}+\dots\nonumber \\
 &= \sum_{\ell\ge1,\, m} \myE_{\ell m} Y^{\ell m}(n^{(i)})\,.
\end{align}
From Eq.~(\ref{abdriftgenobs}) we infer that $\myB(n^{(i)})=0$ and the dipolar components of  $\myE(n^{(i)})$ are
\begin{equation}
\myE_i = \frac{v^{(i)}_{s}-v^{(j)}_{\obs}}{a_{\obs}\chi_{s\obs}}+\dot{v}_{\obs}^{(i)}\,.
\end{equation}
Note that a dipole in $\myB(n^{(i)})$ corresponds to a global
infinitesimal rotation, hence the direction drift generated by our
local velocity cannot be mixed with such effect.

Finally, we note that one could have chosen to decompose the direction drift directly in terms of spin-weighted spherical harmonics; this is actually equivalent to what we are doing since the covariant derivative  on the unit 2-sphere $D^{(i)}$ coincides with the spin-raising operator spherical harmonics~\cite{Goldberg:1966uu,Durrer:2008eom}. 


\subsection{Discussion on velocities}\label{disc-velocities}

The previous sections derive the redshift and direction drifts for both comoving and general observers in a strictly spatially homogeneous and isotropic FL spacetime. Let us have a closer look to our result~(\ref{abdriftgenobs}) concerning the aberration drift. It contains two contributions: the first one is the parallax drift, which encompasses the change of parallax due to the motions of the source and observer, and the second, the aberration drift, is similar to the Bradley aberration. While the first depends on the distance of the source, the second depends only on the acceleration of the observer, a property that we have already explained in our heuristic argument. Other works, such as Refs.~\citep{Rasanen:2013swa,Korzynski:2017nas}, have also computed in more general settings these two contributions to the total variation of direction, employing a different terminology.

To compare the two contributions, we must describe the different motions involved, which comprise the motion of our Local Group of galaxies with respect to the CMB, the motion of the Milky Way with respect to the Local Group, the motion of the Sun around the Milky Way, and finally the motion of the Earth around the Sun. To this purpose, we write
\begin{equation}
v^{(i)}_\obs=v^{(i)}_{\rm{LG}}+v^{(i)}_{\rm{MW}}+v^{(i)}_{\odot}+v^{(i)}_{\oplus}
\end{equation}
and analyze each motion separately. We also provide rough estimates of the order of magnitude of the aberration and parallax drifts, assuming that the motions of the sources are averaged out.

\subsubsection{Local group velocity, $v^{(i)}_{\rm{LG}}$}\label{LG}

Let us start by considering the motion of our Local Group of galaxies. This motion is given by solving the geodesic equation for a point particle with 4-velocity $\tilde{u}^{\mu}$ in a flat FL universe. In the absence of perturbations of the gravitational potential, the solution is simply $\dot{v}^{(i)}_\obs=-H_\obs v^{(i)}_\obs$. The comoving radial distance from the observer to the source satisfies
\begin{equation}
\frac{a_{\obs}\chi_{s\obs}}{D_{H_\obs}}=\int_0^z\frac{\dd z^\prime}{E(z^\prime)},
\end{equation}
where $E(z)=H(z)/H_\obs$. This allows us to rewrite Eq.~(\ref{abdriftgenobs}) as
\begin{equation}
\frac{\delta_{12}\tilde{n}^{(i)}_{\obs}}{\delta t_{\obs}}=\frac{\bot_{j}^{i}}{D_{H_\obs}}\left[\frac{\left(v^{(j)}_{s}-v^{(j)}_{\obs}\right)}{\int_0^z\frac{\dd z^\prime}{E(z^\prime)}}-v_{\obs}^{(j)}\right],
\end{equation}
where $D_{H_\obs}=H_\obs^{-1}$ is the Hubble radius today. Now, $E(z)$ can be evaluated through the Friedmann equations, and in a $\Lambda$CDM scenario it is given by
\begin{equation}
E(z)=\sqrt{\Omega_{m}^\obs(1+z)^3+\Omega_{\Lambda}^\obs},
\end{equation}
where $\Omega_{m}^\obs$ and $\Omega_{\Lambda}^\obs$ are the energy density parameters for matter and dark energy today, respectively. Note that since the parallax drift is redshift dependent through $E(z)$, it can be distinguished from the aberration drift, which does not depend on $z$. To compare their magnitude, we use the fiducial cosmology $\Omega_{m}^\obs=0.31$ and $\Omega_{\Lambda}^\obs=0.69$, following the latest Planck values~\cite{Ade:2015xua}. 

\begin{figure}
\centering
\includegraphics[width=1.01\columnwidth]{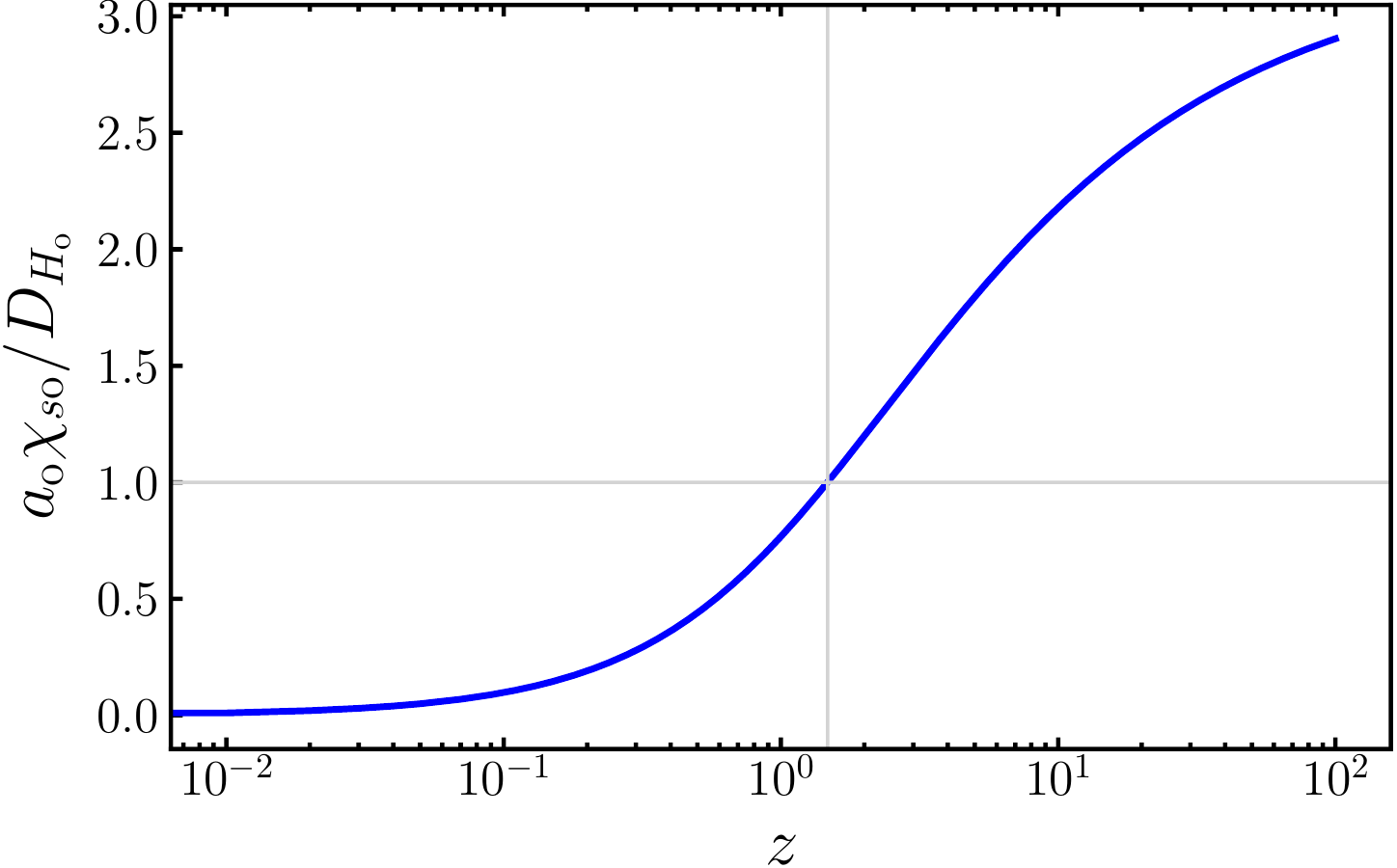}
 \caption{Behavior of the function $\frac{a_{\obs}\chi_{s\obs}}{D_{H_\obs}}=\int_0^z\frac{\dd z^\prime}{E(z^\prime)}$ as a function of $z$. Note that it saturates for large $z$. The vertical line corresponds to $z=1.48$.} 
\label{fig2}
\end{figure}

Assuming comoving sources, the parallax drift is comparable to the aberration drift for $z\approx1.48$, which corresponds to sources  such that $a_{\obs}\chi_{s\obs}\sim D_{H_\obs}$. At smaller redshifts, the parallax drift dominates. Since $1/\int_0^z\frac{\dd z^\prime}{E(z^\prime)}$ saturates\footnote{For large redshifts, $E(z)\approx\sqrt{\Omega_{m}^\obs(1+z)^3}$, such that $1/\int_0^z\frac{\dd z^\prime}{E(z^\prime)}\approx\sqrt{\Omega_{m}^\obs}/2=0.28$. The $0.31$ value mentioned on the text is found when taking into account the nonzero value of $\Omega_{\Lambda}^\obs$.} to approximately $0.31$ at large redshifts (Figure \ref{fig2}), the parallax drift will always contribute to more than $25\%$ of the total signal. This saturation was already noted in Ref.~\citep{Rasanen:2013swa}. Considering the LG has an approximate velocity of $620 \, \text{km} / \text{s}$ with respect to the CMB rest frame \citep{Tully:2007ue}, we expect the aberration drift to be of order $0.03\,\mu \text{as}/\text{yr}$. For sources with comoving radial distance of $D_{H_\obs}$, the parallax drift is of the same magnitude, and remains so even for further sources. For closer sources, e.g., at $0.1D_{H_\obs}$, the parallax drift is more significant, at roughly $0.3\,\mu \text{as}/\text{yr}$.

Our estimates can be compared to Ref.~\cite{2018arXiv180204495B}, and essentially its Eq.~(5). To that purpose we pick up a system of coordinates on the celestial 2-sphere such that
\begin{equation}\label{cordsphe}
\tilde{n}^{(i)}_{\obs}=\left( \sin\tilde{\theta}\cos\tilde{\phi},\sin\tilde{\theta}\sin\tilde{\phi},\cos\tilde{\theta}\right)
\end{equation}
again with the convention that a tilde denotes quantities as seen by the general observer. Following Ref.~\cite{2018arXiv180204495B}, we align the $z$-axis with the velocity of the observer. 
In that coordinates system, the projection of Eq.~(\ref{abdriftgenobs}) on the $z$-axis gives, to first order in velocities,
\begin{equation} \label{eq1}
\begin{split}
\frac{\dd\tilde{\theta}}{\dd t}=&\frac{1}{a_{\obs}\chi_{s\obs}}v_{\obs}\sin{\theta}-\dot{v}_{\obs}\sin{\theta}\\
&-\frac{1}{a_{\obs}\chi_{s\obs}}\left(\frac{v^{(z)}_{s}-\cos{\theta}n_{(i)}v^{(i)}_{s}}{\sin{\theta}}\right).
\end{split}
\end{equation}
Since we are in a pure FL, we have to set $\Phi=\Psi=0$ in the results of Ref.~\cite{2018arXiv180204495B}. Hence, at first order in velocity, the proper time of the general observer and the cosmic time coincide. Comparing the above to Eq. (5) of Ref.~\cite{2018arXiv180204495B}, the aberration contribution to the drift matches with our result. The $\dot{\theta}$ of their Eq. (5) accounts for the parallax drift due to the motions of the source, which can be suppressed by averaging over many sources. Here, this effect is explicit in the last term of Eq.~\eqref{eq1}. The authors of Ref.~\cite{2018arXiv180204495B} are concerned with the aberration drift effect, and argue that the parallactic effects from the motion of the observer can be subtracted from observations. The first term of Eq.~\eqref{eq1} is thus not present in Ref.~\cite{2018arXiv180204495B}. If the effect from the velocities of the sources is averaged out, we are indeed left with the parallax drift due to the motion of the observer and the aberration drift. The total drift thus reads
\begin{equation} \label{eq2}
\frac{\dd\tilde{\theta}}{\dd t}=\frac{1}{a_{\obs}\chi_{s\obs}}v_{\obs}\sin{\theta}-\dot{v}_{\obs}\sin{\theta}.
\end{equation}
Our method of comparing two infinitesimally close geodesics has the advantage of making explicit each of these effects. We show how to apply it in less symmetric spacetime in \S~\ref{sec:Bianchi-I}.

\subsubsection{Milky Way velocity, $v^{(i)}_{\rm{MW}}$}

The main source of acceleration for the Milky Way inside the Local Group is the gravitational potential from M31 (Andromeda). Thus, we approximate the acceleration of the Milky Way by $\dot{v}_{\rm{MW}}=GM_{\rm{M31}}/R^2_{\rm{M31}}$. This gives the magnitude of the aberration drift: with $R_{\rm{M31}}\approx0.8\,\text{Mpc}$ and $M_{\rm{M31}}\approx1.3\times10^{12}\rm{M}_\odot$~\cite{Corbelli:2009nc}, it gives a typical magnitude of  $0.006\,\mu\text{as}/\text{yr}$ for aberration drift.

The contribution of the parallax drift is estimated from $v_{\rm{MW}}/a_\obs \chi_{s\obs}$. The Milky Way's velocity with respect to the local group is estimated to be of order $135\,\text{km}/\text{s}$ \citep{Tully:2007ue}. For sources at a distance $a_\obs \chi_{s\obs}=D_{H_\obs}$, the parallax drift for is then  of the order $0.006\,\mu\text{as}/\text{yr}$. For sources further than $a_\obs \chi_{s\obs}\gtrsim D_{H_\obs}$, the total direction drift for the motion of the MW inside the LG is roughly one order of magnitude smaller than the drift of the LG with respect to the CMB.

\subsubsection{Sun velocity, $v^{(i)}_{\odot}$}

Let us estimate the direction drift due to the motion of the Sun around the Galactic Center. From Kepler's laws, the acceleration of the Sun around the Galactic Center is  $\dot{v}_{\odot}=v_{\odot}^2/R_{\rm{GC}}$, where $R_{\rm{GC}}$ is the distance from the Sun to the Galactic Center. The parallax drift is simply estimated from $v_{\odot}/a_\obs \chi_{s\obs}$. Considering $v_{\odot}\approx230 \text{km/s}$ \citep{Tully:2007ue} and the distance to the Galactic Center to be $R_{\rm{GC}}\approx8\text{kpc}$, the aberration drift and the parallax drift are found to be of the same order for sources such that $a_\obs \chi_{s\obs}\simeq cR_{\rm{GC}}/v_{\odot} \approx 10\text{Mpc}$. This shows that both the parallax and aberration drifts may contribute significantly to the total signal. The aberration drift is of order $4\,\mu\text{as}/\text{yr}$ and, at $a_\obs \chi_{s\obs}=D_{H_\obs}$, the parallax drift is of order $0.01\,\mu\text{as}/\text{yr}$. Thus, the aberration drift contribution from $v_{\odot}$ is $100$ times larger than the aberration drift due to the motion of the LG.

\subsubsection{Earth velocity, $v^{(i)}_{\oplus}$}
 
The direction drift contribution from the motion of the Earth around the Sun comprises an annual modulation to the total signal which is not cumulative. Since the motion of the Earth around the Sun is well understood, it is expected that this signal can be subtracted.

\subsubsection{Final remarks on velocities}

Table \ref{Table1} summarizes the previous estimations of the aberration and parallax drifts. They are denoted $\dot{v}$ and $v/a_\obs\chi_{s\obs}$, respectively. We also consider different comoving radial distances to the source for the parallax drift: $3D_{H_\obs}$, which corresponds to $z\gtrsim100$, at which $a_\obs \chi_{s\obs}$ saturates; $D_{H_\obs}$, which corresponds to $z\approx1.48$; and $0.1D_{H_\obs}$, which corresponds to $z\approx0.1$.

Quasars are found in a range of redshifts up to $z=5$, but are mostly around $z=1$ and $z=2$ \citep{2012A&A...543A.100R}. This means that for both the Local Group and the Milky Way motions, the aberration and parallax drifts are of the same order of magnitude.

We must finally note that our estimations completely ignore the direction of the different velocities, which must be taken into account for a precise analysis.

\begin{table}[h!]
  \begin{center}
    \begin{tabular}{|c
    				|S[table-figures-decimal=1]
    				|S[table-figures-decimal=1]
    				|S[table-figures-decimal=1]|}
\hline
      \, & LG & MW & $\odot$ \\
      \hline
      $\dot{v}$ & 0.03 & 0.006 & 4.0\\
      $v/(3D_{H_\obs})$ & 0.001 & 0.002 & 0.003\\
      $v/D_{H_\obs}$ & 0.03 & 0.006 & 0.01\\
      $v/(0.1D_{H_\obs})$ & 0.3 & 0.06 & 0.1\\
\hline
    \end{tabular}
    \caption{Estimation of aberration and parallax drifts for different motions. The values are expressed in units of $\mu\text{as}/\text{yr}$.} \label{Table1}
  \end{center}
\end{table}

\section{Perturbed FL spacetimes}\label{Sec3b}

We extend the previous analysis to take into account the effects of the large scale structure. The universe is then described by a perturbed FL spacetime with geometry
\begin{equation}\label{perFL}
 \dd s^{2}=a^{2}\left[-\left(1+2\Phi\right)\dd\eta^{2}+\left(\delta_{ij}+h_{ij}\right)\dd x^{i}\dd x^{j}\right],
\end{equation}
with
\begin{equation}
h_{ij}=-2\Psi\delta_{ij}+2\partial_{(i}E_{j)}+2E_{ij}\,.
\end{equation}
This defines our choice of Newtonian gauge where the scalar modes are described by the two gravitational potentials, $\Phi$ and $\Psi$, vector perturbations are described by a transverse vector $E_i$ ($\partial_i E^i=0$) and $E_{ij}$ represents the traceless and transverse tensor perturbation ($E^i_i=0=\partial_i E^{ij}$) describing gravitational waves.
 
The gravitational potentials have two effects on the direction drift. The first is a direct effect on the drift while the second is an effect on the motion of the observer. More precisely, it will affect the geodesic motion of the Local Group with respect to the CMB since then the geodesic equation for a point particle with 4-velocity $\tilde{u}^\mu$ is
\begin{equation}
\dot{v}^{(i)}_\obs=-H_\obs v^{(i)}_\obs-\partial^i \Phi_\obs.
\end{equation}
Ref.~\cite{2018arXiv180204495B} describes how measurements of the cosmological aberration drift could be used to evaluate the contribution from $\partial^i\Phi_\obs$. In fact, neglecting decaying and vorticity modes, one can show that the contribution from the gravitational potential is proportional to $H_\obs v^{(i)}_\obs$, with the proportionality factor being given by a model-dependent parameter. In standard GR, this parameter is of order one, and the contribution of $\partial^i\Phi_\obs$ is of the same order of magnitude as the cosmological aberration drift~\cite{2018arXiv180204495B}.

\subsection{Null geodesics}

The study of the null geodesics is simpler once one uses the standard trick that they are conformally invariant. The metric~(\ref{perFL}) can be rescaled as $\dd s^{2}=g_{\mu \nu}\dd x^{\mu}\dd x^{\nu}=a^{2}\hat{g}_{\mu \nu}\dd x^{\mu}\dd x^{\nu}$. Then, if $k^{\mu}$ is the tangent vector to a null geodesic of $g_{\mu \nu}$, then $\hat{k}^{\mu}=a^{2}k^{\mu}$ is the tangent vector to a null geodesic of $\hat{g}_{\mu \nu}$, where from now on the overhat denotes quantities evaluated in the conformal space.

The conformal null geodesic vector $\hat{k}^{\mu}$ can be decomposed as 
\begin{equation}
\hat{k}^{\mu}=\hat{\omega}\left(1,n^{i}\right).
\end{equation}
We define $\hat{\bar{\omega}}_\obs$  and $\bar{n}^{i}$ the (constant) background values of $\hat{\omega}$ and  and $n^{i}$, respectively. To first order in perturbations, the $0$-component of the geodesic equation is
\begin{equation} \label{0geodpertfl}
\frac{1}{\hat{\omega}}\frac{\dd\hat{\omega}}{\dd\eta}=\Phi^{\prime}-2\frac{\dd\Phi}{\dd\eta}-\frac{1}{2}\bar n^{i}\bar n^{j}h^{\prime}_{ij}.
\end{equation}
where we used that $\frac{\dd}{\dd\eta} \equiv \frac{1}{\hat{\omega}}\hat{k}^{\mu}\partial_{\mu}=\frac{\partial}{\partial \eta}+n^{i}\partial_{i}$ and the prime denotes the partial derivative with respect to $\eta$. It can be integrated to give
\begin{equation} \label{omegapertfl}
\begin{split}
\hat{\omega}_{s}-\hat{\omega}_{\obs}&=\hat{\bar{\omega}}_{\obs}\left[-2\left(\Phi_{s}-\Phi_{\obs}\right) \right.\\
&\quad\left.+\int_{\eta_{\obs}}^{\eta_{s}}\left(\Phi^{\prime}-\frac{1}{2}\bar{n}^{i}\bar{n}^{j}h^{\prime}_{ij}\right)\dd\eta\right].
\end{split}
\end{equation}
The energy of the photon requires  the 4-velocity of a comoving observer in perturbed FL universe to be computed. But, as long as we are dealing with first order effects, we can ignore the spatial part of the 4-velocity and then include it as described in the previous section. So, in conformal time,
\begin{equation}
\hat{u}_{\mu}=(-1-\Phi,\bm{0}).
\end{equation}
Since $\hat{u}_{\mu}=\frac{1}{a}u_{\mu}$, we deduce that the energy of the photon as measured by an observer with 4-velocity $u^{\mu}$ is
\begin{align}
\omega &= -k^{\mu} u_{\mu}= \frac{\hat{\omega}}{a}\left(1+\Phi\right),
\end{align}
from which we deduce the redshift
\begin{align}
1+z&=\frac{\omega_{s}}{\omega_{\obs}}=\frac{a_{\obs}}{a_{s}}\frac{\hat{\omega}_{s}}{\hat{\omega}_{\obs}}\left[1+\left(\Phi_{s}-\Phi_{\obs}\right)\right].
\end{align}
Now, $\hat{\omega}_{s}/\hat{\omega}_{\obs}$ can be obtained from Eq.~\eqref{omegapertfl} so that
\begin{equation} \label{rsdpertfl}
\begin{split}
1+z&=\frac{a_{\obs}}{a_{s}}\left[1-\left(\Phi_{s}-\Phi_{\obs}\right)\right.\\
&\quad\left.+\int_{\eta_{\obs}}^{\eta_{s}}\left(\Phi^{\prime}-\frac{1}{2}\bar{n}^{i}\bar{n}^{j}h^{\prime}_{ij}\right)\dd\eta\right].
\end{split}
\end{equation}
It decomposes in scalar, vector and tensor contributions as $1+\bar z + \delta z^{(S)} + \delta z^{(V)} +\delta z^{(T)}$, with
\begin{subequations}
\begin{align}
\delta z^{(S)} =& \frac{a_{\obs}}{a_{s}}\left[-(\Phi_s-\Phi_\obs) + \int_{\eta_{\obs}}^{\eta_{s}}\Theta'\dd\eta\right]\,, \\
\delta z^{(V)} &= -\frac{a_{\obs}}{a_{s}}\int_{\eta_{\obs}}^{\eta_{s}} \partial_{(i}E'_{j)}\bar{n}^{i}\bar{n}^{j}\,\dd\eta\,, \\
\delta z^{(T)} &= -\frac{a_{\obs}}{a_{s}}\int_{\eta_{\obs}}^{\eta_{s}} E'_{ij}\bar{n}^{i}\bar{n}^{j}\,\dd\eta\,, \label{perturbedrs}
\end{align}
\end{subequations}
where we have defined the standard lensing potential as
\begin{equation}
 \Theta = \Phi + \Psi.
\end{equation}

\subsection{Redshift drift}

Evaluating the redshift drift is more involved in a perturbed FL universe since the constant time hypersurfaces are no longer homogeneous. Considering a second geodesic corresponding to an observation at $t_\obs+\delta t_\obs$, the change in redshift $\delta_{12} z$ is obtained from Eq.~\eqref{rsdpertfl} to be
\begin{equation} \label{deltazpertfl}
\frac{\delta_{12} z}{1+z}=\left(H_{\obs}\delta t_{\obs}-H_{s}\delta t_{s}\right)+\delta_{12}\Upsilon
\end{equation}
where 
\[\Upsilon=-\left(\Phi_{s}-\Phi_{\obs}\right)+\int_{\eta_{\obs}}^{\eta_{s}}\left(\Phi^{\prime}-\frac{1}{2}\bar{n}^{i}\bar{n}^{j}h^{\prime}_{ij}\right)\dd\eta\] 
and $\delta_{12}\Upsilon$ stands for the difference of $\Upsilon$ between its value on the second and first geodesics. To evaluate the drift of the integrated terms in $\Upsilon$, one has to take into account how they vary spatially from one geodesic to the next as space is no longer homogeneous. This calculation is detailed in Appendix \ref{apendixB}, thus yielding
\begin{widetext}
\begin{align}
 \delta_{12}\Upsilon&=-\left(\dot{\Phi}_{s}\delta t_{s}-\dot{\Phi}_{\obs}\delta t_{\obs}\right)  \label{intterms}+\int_{\eta_{\obs}}^{\eta_{s}}\left(\Phi^{\prime \prime}-\frac{1}{2}\bar{n}^{i}\bar{n}^{j}h^{\prime \prime}_{ij}\right)\, \dd\eta \,  \delta\eta_\obs. \nonumber
\end{align}
Plugging this back into Eq.~\eqref{deltazpertfl}, we get
\begin{align}
&\frac{\delta_{12} z}{1+z}=\left(H_{\obs}\delta t_{\obs}-H_{s}\delta t_{s}\right)-\left(\dot{\Phi}_{s}\delta t_{s}-\dot{\Phi}_{\obs}\delta t_{\obs}\right)+\int_{\eta_{\obs}}^{\eta_{s}}\left(\Phi^{\prime \prime}-\frac{1}{2}\bar{n}^{i}\bar{n}^{j}h^{\prime \prime}_{ij}\right) \dd\eta\, \delta\eta_\obs.
\end{align}
To obtain the redshift drift, we must also take into account the difference between the observer's proper time $\tau$ and the cosmic time $t$, $\delta \tau=(1+\Phi)\delta t$, and then use Eq.~\eqref{tzrelation} to conclude that
\begin{equation} \label{zdotpertfl}
\frac{\delta_{12} z}{\delta \tau_{\obs}}=(1+z)\left[H_{\obs}(1-\Phi_{\obs})+\dot{\Phi}_{\obs}\right]-\left[H_{s}(1-\Phi_{s})+\dot{\Phi}_{s}\right]+\frac{(1+z)}{a_{\obs}}\int_{\eta_{\obs}}^{\eta_{s}}\left(\Phi^{\prime \prime}-\frac{1}{2}\bar{n}^{i}\bar{n}^{j}h^{\prime \prime}_{ij}\right) \dd\eta\,.
\end{equation}
\end{widetext}

\subsection{Direction drift}

To evaluate the direction drift, we start with the $i$-component of the geodesic equation,
\begin{align}
\frac{\dd n^{i}}{\dd \eta}&-\bar{n}^{i}\frac{\dd \Phi}{\dd \eta}+\underline\bot^{i}_{j}\partial^{j}\Phi-\frac{1}{2}\bot^{i}_{j}\partial^{j}(\bar{n}^k \bar{n}^l h_{kl}) \nonumber \\
&+\bar{n}^{j}\frac{\dd h^{i}_{j}}{\dd \eta}-\frac{1}{2}\bar{n}^{i}\bar{n}^{j}\bar{n}^{k}\frac{\dd h_{jk}}{\dd \eta}=0.
 \label{igeodpertbfl}
\end{align}
With the tetrad for the conformal space defined by
\begin{equation}
\bm{\epsilon}_{(0)}=\left(1-\Phi\right)\partial_{\eta}, \quad
\bm{\epsilon}_{(i)}=\left(\delta_{i}^{j}-\frac{1}{2}h_{i}^{j}\right)\partial_{j},
\end{equation}
the direction vector is decomposed in tetrad components as
\begin{equation}
n^{(i)}=\frac{\hat{k}^{(i)}}{\hat{k}^{(0)}},
\end{equation}
so that
\begin{equation} \label{eini}
n^{i}=n^{(i)}+\Phi n^{(i)}-\frac{1}{2}h^{i}_{j}n^{(j)}.
\end{equation}
Plugging this decomposition into Eq.~\eqref{igeodpertbfl} gives
\begin{align}
\frac{\dd n^{(i)}}{\dd \eta}=&-\underline\bot^{i}_{j}\partial^{j}\Phi+\frac{1}{2}\underline\bot^{i}_{j}\partial^{j}\left[\bar{n}^{(k)} \bar{n}^{(l)} h_{kl}\right]
\nonumber \\
&-\frac{1}{2}\underline\bot^{i}_{j}\bar{n}^{(k)}\frac{\dd h^{j}_{k}}{\dd \eta}\,.
\end{align}
By integrating this equation and then replacing $n^{(i)}$ by $n^i=\frac{\dd x^i}{\dd \eta}$ using Eq.~\eqref{eini}, one gets the null geodesic equation,
\begin{align}
\frac{\dd x^i}{\dd \eta}&=n^{(i)}_\obs+\bar{n}^{(i)}_\obs \Phi-\frac{1}{2}\bar{n}^{(j)}_\obs h^i_j\\
&-\frac{1}{2}\underline\bot^i_j\bar{n}^{(k)}_\obs(h^j_k-h^j_{\obs k})\nonumber \\
&-\underline\bot^i_j\partial^j \int_{\eta_\obs}^{\eta}\left\lbrace \Phi - \frac{1}{2}\left[\bar{n}^{(k)}_\obs\bar{n}^{(l)}_\obs h_{kl}\right]\right\rbrace \dd \eta.\nonumber
\end{align}
After integration from $\eta_\obs$ to $\eta_s$, its scalar, vector and tensor parts are 
\begin{widetext}
\begin{align}\label{geodeqpertbfl}
& x^i_s-x^i_\obs =(\eta_s-\eta_\obs)n^{(i)}_\obs \nonumber\\
&+\bar{n}^{(i)}_\obs\int_{\eta_\obs}^{\eta_s}\Theta \, \dd\eta -\underline\bot^i_j\int_{\eta_\obs}^{\eta_s}(\eta_s-\eta)\partial^j\Theta\,\dd \eta \\
& +(\eta_s-\eta_\obs)\underline\bot^i_j\bar{n}^{(k)}_\obs \partial_{(k}E^{j)}_{\obs} -\underline\bot^i_j\bar{n}^{(k)}_\obs\int^{\eta_s}_{\eta_\obs}\partial_{(k}E^{j)}\,\dd\eta-\bar{n}^{(j)}_\obs\int_{\eta_\obs}^{\eta_s}\!\!\!\partial_{(j}E^{i)}\,\dd\eta +\underline\bot^i_j \int_{\eta_\obs}^{\eta_s}(\eta_s-\eta) \bar{n}^{(k)}_\obs\bar{n}^{(l)}_\obs \partial^j  \partial_{k}E_{l}\,\dd\eta \nonumber \\
& +(\eta_s-\eta_\obs)\underline\bot^i_j\bar{n}^{(k)}_\obs E^j_{\obs k} -\underline\bot^i_j\bar{n}^{(k)}_\obs\int^{\eta_s}_{\eta_\obs}E^j_k\,\dd\eta-\bar{n}^{(j)}_\obs\int_{\eta_\obs}^{\eta_s}\!\!\!E^i_j\,\dd\eta +\underline\bot^i_j \int_{\eta_\obs}^{\eta_s}(\eta_s-\eta) \bar{n}^{(k)}_\obs\bar{n}^{(l)}_\obs \partial^j  E_{kl}\,\dd\eta \nonumber\,,
\end{align}
\end{widetext}
where we have performed an integration by parts to express the double integral as single integrals. We shall now write this equation for a second geodesic. The terms evaluated at fixed endpoints are simple to compute for the second geodesic. For the integrated terms,we refer to Appendix \ref{apendixB}, where the calculation is presented in detail. We need to take into account the difference between proper time of the observer and cosmic time, and also make use of Eqs.~\eqref{rsdpertfl} and~\eqref{tzrelation}. Splitting in scalar, vector and tensor modes, the direction drift is finally given by
\begin{widetext}
\begin{subequations}
\begin{align}
\frac{\delta_{12} n^{(i)}_\obs}{\delta\tau_\obs}^{(S)} & =-\frac{\underline \bot^i_j }{a_\obs}\int_{\eta_\obs}^{\eta_s}\frac{(\eta_s-\eta)}{\chi_{s\obs}}\partial^j\Theta'\dd \eta\,, \\
\frac{\delta_{12} n^{(i)}_\obs}{\delta \tau_\obs}^{(V)}&=-\frac{1}{a_\obs}\underline\bot^i_j \bar{n}^{(k)}_\obs \partial_{(k}E^{\prime j)}_{\obs}-\frac{2\underline\bot^i_j\bar{n}^{(k)}_\obs}{a_\obs\chi_{s\obs}}\int_{\eta_\obs}^{\eta_s}\!\!\!\partial_{(k}E^{\prime j)}\dd\eta +\frac{\underline\bot^i_j}{a_\obs}
\int_{\eta_\obs}^{\eta_s}\frac{(\eta_s-\eta)}{\chi_{s\obs}} \bar{n}^{(k)}_\obs\bar{n}^{(l)}_\obs \partial^j\partial_{(k} E^{\prime}_{l)}\,\dd\eta\,,\\
\frac{\delta_{12} n^{(i)}_\obs}{\delta \tau_\obs}^{(T)}&=-\frac{1}{a_\obs}\underline\bot^i_j \bar{n}^{(k)}_\obs E^{\prime j}_{\obs k}-\frac{2\underline\bot^i_j\bar{n}^{(k)}_\obs}{a_\obs\chi_{s\obs}}\int_{\eta_\obs}^{\eta_s}\!\!\!E^{\prime j}_k\dd\eta+\frac{\underline\bot^i_j}{a_\obs}
\int_{\eta_\obs}^{\eta_s}\frac{(\eta_s-\eta)}{\chi_{s\obs}} \bar{n}^{(k)}_\obs\bar{n}^{(l)}_\obs \partial^j E^{\prime}_{kl}\,\dd\eta.\label{3.56}
\end{align}
\end{subequations}
\end{widetext}

\subsection{Summary}

This section provides the first derivation of the direction drift and redshift drift in a perturbed FL universe. As such each has three contributions arising respectively from scalar, vector and tensor modes. First, the scalar part of our expression~(\ref{zdotpertfl}) corrects a mistake in the only expression proposed in the literature so far and first published in Ref.~\cite{Uzan:2007tc}. Such an expression plays an important role in estimating the expected cosmological variance of the redshift drift. Note that the scalar mode contribution has to be combined with Eq.~(\ref{abdriftgenobs}) to include the effect of the motions of the observer and the sources.

Concerning gravity waves, several results~\cite{1975GReGr...6..439E,wjk,Book:2010pf,Pyne:1995iy,Gwinn:1996gv,Moore:2017ity} have been used, in particular by Pulsar Timing Array experiments. The result of Ref.~\cite{Book:2010pf} gives the perturbed values of $z$ and $n^{(i)}$ with respect to the background $\bar{z}$ and $\bar{n}^{(i)}$ values. Their equation (28) is directly comparable to Eq.~\eqref{perturbedrs}, and the results match considering the relationship between the parameters $\lambda$ and $\eta$ and that Eq. (28) is for a pure Minkowski spacetime. To compute the perturbation of $n^{(i)}$ in our framework, we start from Eq.~\eqref{geodeqpertbfl} and split it into its background and perturbed values. Noticing that the positions of the source and observer are fixed in the ``straight geodesic" approximation, we take the perpendicular projection of Eq.~\eqref{geodeqpertbfl} to find
\begin{align}
&\delta n^{(i)}=-\underline\bot^i_j \bar{n}^{(k)}E^j_{\obs k}-\frac{2\underline\bot^i_j \bar{n}^{(k)}}{\chi_{s\obs}}\int_{\eta_\obs}^{\eta_s}E^j_k\,\dd\eta \nonumber\\
&+\underline\bot^i_j\int_{\eta_\obs}^{\eta_s}\frac{\left(\eta_s-\eta\right)}{\chi_{s\obs}} \bar{n}^{(k)}_\obs \bar{n}^{(l)}_\obs \partial^j E_{kl} \, \dd\eta.
\end{align}
This matches with the Eq. (56) of Ref.~\cite{Book:2010pf} if we take into account our opposite sign in defining the direction vector and again, the relationship between the parameters $\zeta$ and $\eta$. Thus, the plane wave expansion used in Refs.~\cite{1975GReGr...6..439E,wjk,Book:2010pf,Pyne:1995iy,Gwinn:1996gv,Moore:2017ity} is compatible with our analysis. Indeed their results only provide deviation of $z$ and $n^{(i)}$ from their background value and do not provide their drifts computed here.

\section{Spatially homogeneous and anisotropic spacetimes: Bianchi I case} \label{sec:Bianchi-I}

Bianchi I spacetime is one of the solutions of a class of spatially anisotropic and homogeneous spacetimes (see Ref.~\cite{Ellis1969} for details). Being homogeneous, these spaces still enjoy three Killing vectors associated with the three spatial translations. The spatial sections are also Euclidean, and the spacetime metric is given by
\begin{equation} \label{bImetric}
 \dd s^2=-\dd t^{2}+a^2(t)\sum_{i,j} e^{2\beta_{i}(t)}\delta_{ij}\dd x^{i}\dd x^{j}\,,
\end{equation}
where the $\beta_{i}$ are three directional scale factors which satisfy $\sum_{i}\beta_{i}=0$, and $a$ is the average scale factor defined by the volume 
expansion.

\subsection{Null geodesics}

Thanks to these  Killing vectors, we still have
 \[k_{i}=\text{cte}\] 
along any null geodesic for a photon with 4-momentum $k^{\mu}=\omega(u^{\mu}-n^{\mu})$, as long as we use the Cartesian coordinates introduced in Eq.~\eqref{bImetric}. It leads to
\begin{equation} \label{eq:geod1}
a^{2}e^{2\beta_{i}}k^{i}=\text{cte}\,,
\end{equation}
with no summation over $i$. The worldline of a photon is again given  
\[\frac{\dd x^{i}}{\dd\lambda}=k^{i},\] 
where $\lambda$ is the parameter along the geodesic. In conformal time, the orthonormal tetrads are explicitly given by
\begin{equation}
 \bm{e}_{(0)}=\partial_{t}, \quad 
 \bm{e}_{(i)}=a^{-1}e^{-\beta_{i}}\partial_{i},
\end{equation} 
in terms of which Eq. (\ref{eq:geod1}) becomes
\begin{equation} \label{eq:geod2}
a\omega e^{2\beta_{i}}\frac{\dd x^{i}}{\dd\eta}=-a_{\obs}\omega_{\obs}e^{\beta_{i}^{0}}n_{\obs}^{(i)}
\end{equation}
where $n_{\obs}^{(i)}$ are the tetrad basis components of $n_{\obs}^{i}$. We also define the ``conformal'' shear by
\begin{equation}
 \sigma_{ij}=\beta'_i\hbox{e}^{2\beta_i}\delta_{ij},
\end{equation}
which is a symmetric trace-free tensor.

\subsection{Direction and redshift drifts to first order in shear}

First, we note that the 0-component of the geodesic takes the form
\begin{equation} \label{fullgeodeq}
\frac{1}{a\omega}\frac{\dd(a\omega)}{\dd\eta}+\beta^{\prime}_{i}n^{(i)}n_{(i)}=0\,,
\end{equation} 
a solution of which is
\begin{equation} \label{fullexp1}
a\omega=a_{\obs}\omega_{\obs}\exp\left(-\int^{\eta}_{\eta_{\obs}}\beta^{\prime}_{i}n^{(i)}n_{(i)}\dd\eta\right)\,.
\end{equation}
Note that we omit the sums involving $\beta_{i}$ for simplicity. To avoid confusion, one should keep in mind that an expression like $\beta_{i}n^{(i)}$ is not summed, while $\beta_{i}n^{(i)}n_{(i)}$ is. 

We shall now perform a small shear approximation and consider only the lowest order terms in $\sigma_{ij}$ or, equivalently, in $\beta_i'$. In this limit Eq.~\eqref{fullexp1} becomes
\begin{equation} \label{omegasmallb}
\frac{\omega}{\omega_{\obs}}\simeq\frac{\bar{\omega}}{\bar{\omega}_{\obs}}\left[1-\left(\beta_{i}-\beta_{i}^{\obs}\right)\bar{n}^{(i)}\bar{n}_{(i)}\right],
\end{equation}
where an overbar denotes a FL value since the Bianchi I spacetime can be thought as a homogeneous perturbation of the FL spacetime. Since $\bar{n}^{(i)}$ is constant, the small shear approximation is related to a ``straight geodesic" approximation, or to the more usual Born approximation in lensing. It follows that Eq.~(\ref{eq:geod2}) becomes
\begin{equation} \label{geodeqsol1}
\begin{split}
x_{s}^{i}-x_{\obs}^{i}&\simeq-\int_{\eta_{\obs}}^{\eta_{s}}\left(1-\beta_{i}\right)\, n^{(i)}_{\obs} \dd\eta \\ 
&+\int_{\eta_{\obs}}^{\eta_{s}}\underline\bot_{j}^{i}\left(\beta_{j}-\beta_{j}^{\obs}\right)\bar{n}^{(j)}\,\dd\eta 
\end{split}
\end{equation} 
where $\underline\bot^{i}_{j}=\delta_{j}^{i}-\bar{n}^{(i)}\bar{n}_{(j)}$ is the perpendicular projector with respect to $\bar{n}^{(i)}$.
Now, we just have to follow the same procedure, i.e., integrating the equations for two nearby geodesics and determine the relation between $\delta \eta_{s}$ and $\delta \eta_{\obs}$. In a Bianchi I, the latter is given by Eq.~\eqref{tzrelation} to be
\begin{equation} \label{fullexp3}
\frac{\delta \eta_{s}}{\delta \eta_{\obs}}=\left[1+\left(\beta_{i}-\beta_{i}^{\obs}\right)\bar{n}^{(i)}\bar{n}_{(i)}\right].
\end{equation}
Hence, we have all the pieces to determine the drifts.

\subsubsection{Direction drift}

Evaluating Eq.~\eqref{geodeqsol1} at $\eta_{s,\obs}$ and $\eta_{s,\obs}+\delta\eta_{s,\obs}$ and using  Eq.~\eqref{fullexp3}, the direction drift is
\begin{equation} \label{abdrift}
\frac{\delta_{12} n^{(i)}}{\delta t_{\obs}}=-\frac{2}{a_{\obs}\chi_{s\obs}}\underline\bot_{j}^{i}\left(\beta_{j}^{s}-\beta_{j}^{\obs}\right)\bar{n}^{(j)}-\underline\bot_{j}^{i}\dot{\beta}_{j}^{\obs}\bar{n}^{(j)},
\end{equation}
where $\chi_{s\obs}=\eta_{\obs}-\eta_{s}$ is the observer-source radial distance in the FL background space. Notice that this expression is similar to the one for a general observers in a FL spacetime, containing a parallax type (the first term above) and an aberration type (the second term) contribution. Note also that the first term is a redshift-dependent one, whereas the second is redshift-independent.

\subsubsection{Redshift drift}

The redshift drift is calculated in a similar way, from Eq.~\eqref{omegasmallb}, to be
 \[
1+z = \frac{\omega_{s}}{\omega_{\obs}} = \frac{a_{\obs}}{a_{s}}\left[1-\left(\beta_{i}^{s}-\beta_{i}^{\obs}\right)\bar{n}^{(i)}\bar{n}_{(i)}\right] 
\]
As in the previous sections, it follows that
\begin{equation} \label{zdrift}
\begin{split}
\frac{\delta_{12} z}{\delta t_{\obs}} &= (1+z)\left(H_{\obs}+\dot{\beta}_{i}^{\obs}\bar{n}^{(i)}\bar{n}_{(i)}\right) \\\ &\qquad \qquad - \left(H_{s}+\dot{\beta}_{i}^{s}\bar{n}^{(i)}\bar{n}_{(i)}
\right)\!,
\end{split}
\end{equation}
in agreement with the analysis of Ref.~\cite{Fleury:2014rea}.

\subsection{Decomposition in multipoles}

As in \S~\ref{SecMultipoles1}, we can decompose the redshift and direction drifts in terms of multipoles, the decomposition being made with respect to $\bar{n}^{(i)}$. Defining
\begin{equation}
{\cal B}_{ij} \equiv \delta_{ij} {\beta}_{i}\qquad 
\dot{\cal B}_{ij} \equiv \delta_{ij} \dot{\beta}_{i}
\end{equation} 
we read from Eq. (\ref{zdrift}) that the Bianchi I spacetime brings a quadrupolar structure to the redshift drift,
\begin{eqnarray}
\myZ_{ij} = (1+z) \dot{\cal B}^{\obs}_{ij}-\dot{\cal B}^{s}_{ij}\,.
\end{eqnarray}
Note however that it adds no dipolar contribution. It is thus in principle distinguishable from the peculiar velocity of the observer in a FL spacetime.

The direction drift in  Bianchi I structure also inherits a quadrupolar contribution in $\myE(\bar{n}^{(i)})$, but not in $\myB(\bar{n}^{(i)})$. From Eq. (\ref{abdrift}), it is
\begin{equation}
\myE_{ij} = -\frac{1}{a_{\obs}\chi_{s\obs}}({\cal B}_{ij}^s -{\cal B}_{ij}^{\obs}) -\frac{1}{2}\dot{\cal B}_{ij}^{\obs}\,.
\end{equation}

\subsection{Discussion}

Equations~(\ref{abdrift}) and~(\ref{zdrift}) provide the expressions of the direction and redshift drifts in a Bianchi I spacetime, in small shear approximation. Indeed  when $\beta_i=0$ for all $i$, we recover the FL expressions for a comoving observer.

Moreover, since the infinite wavelength limit of a gravitational wave is equivalent to a homogeneous shear (i.e., $E_{ij}\rightarrow \beta_i(t)\delta_{ij}$),  it can be checked that the tensor part of Eq.~(\ref{zdotpertfl}) is equivalent to Eq.~\eqref{zdrift} for the redshift drift and that Eq.~(\ref{3.56}) is equivalent to Eq.~(\ref{abdrift}) for the aberration drift.


Let us now compare to existing results, and in particular Refs.~\cite{Quercellini:2010zr,2009PhRvD..80f3527Q} that assume a ``straight geodesic approximation" and in which the results are obtained through a time derivative of the aberration (as defined here). Defining a spatial orthogonal coordinates system aligned with the principal axis of expansion so that, with the notation of Ref.~\cite{Quercellini:2010zr}, ${H_X/H-1= \dot{\beta}_1/H=\Sigma_X}$, etc. and decomposing the direction of observation as in Eq.~(\ref{cordsphe}), it is easy to show that Eqs.~(33-34) of Ref.~\cite{Quercellini:2010zr} (or Eqs.~(6-7) of Ref.~\cite{2009PhRvD..80f3527Q}) are equivalent to $-\underline\bot_{j}^{i}\dot{\beta}_{j}^{\obs}\bar{n}^{(j)}$ up to an overall minus sign. This is only the aberration drift effect of Eq.~(\ref{abdrift}) and these expressions do not contain the parallax drift included in our expression. Indeed,  it cannot be obtained by deriving the expression of the aberration of angles with respect to time since it arises from the fact that the end points of the two geodesics are different.

{Lastly, we can give some crude estimates of the level to which measurements of the direction drift can constrain the shear. For simplicity let us write $\dot{\beta}_\obs$ as a fraction of $H_\obs$, i.e., $\dot{\beta}_\obs\sim\epsilon H_\obs$. Then, we can separate the constraints in two types: early and late anisotropies. For the first type, CMB data severely constrains the value of the shear today to no larger than the observed CMB quadrupole, i.e., $\epsilon < 10^{-5}$. This would lead to an aberration drift of the order of $10^{-4}\,\mu\text{as/yr}$ or smaller. As has been pointed out, this is three order of magnitudes smaller than the expected peculiar velocities in the standard (FL) model, so very unlike to be detected with future experiments~\cite{2009PhRvD..80l3515F}. Late type anisotropies are however more promising. Future surveys of weak-lensing shear such as Euclid could constrain the late anisotropy of the cosmic flow to order $\dot{\beta}_\obs/H_\obs=1\%$~\cite{Pereira:2015jya}. This would imply a signal in the aberration drift of the order $0.1\,\mu\text{as/yr}$ per source, or $1.0\,\mu\text{as}$ over ten years. This is compatible with the figure of $0.4\mu\text{as}$ in ten years coming from the Local Group proper motion with respect to the CMB frame found in Ref.~\cite{2018arXiv180204495B}. In fact, as we pointed out in Section~\ref{LG}, assuming $v_{\text{LG}}\sim 620\,\text{km/s}$ leads to a drift of the order of $0.3\,\mu\text{as}$ in ten years. Moreover, since the shear contribution to the drift is coming from a quadrupole, it can in princible be distinguished from the velocity contribution.}

\section{Conclusions} \label{Sec5}

Our analysis proposes a general method to compute the direction and redshift drifts, making clear the importance of considering two nearby geodescis connecting 
the observer and the source. Two effects have to be considered, the first related to the integral along the line of sight and the second related to the end points of the geodesics. In a non-static spacetime or for non-inertial observer and sources the lapses connecting the geodesics at the source and at the observer differ.

This general method allowed us to first recover the standard formula for the direction and redshift drifts for a general observer in a FL spacetime. The multipolar decomposition shows that for both the redshift and the direction drifts, the observer's velocity only induces a dipole.  We were able to separate the contribution of the velocity to the direction drift into two effects, the parallax drift (which is $z$-dependent) and aberration drift. Comparing the two for various combined motions, we showed that the two do contribute significantly to the total drift. However, the parallax drift is usually considered as a noise to be removed, which can be accomplished using the $z$-dependence. The drift caused by the motion of the Sun around the galactic center must also be removed if one is interested in the cosmological aberration drift.

We also provided an expression for the direction and redshift drifts in a perturbed FL universe for scalar, vector and tensor perturbations. Concerning the scalar part of the redshift drift, we corrected a mistake in the literature. The tensor contribution to both the redshift and direction vector were compared to existing results for a plane (gravitational) wave. We have left a full multipolar analysis and calculation of correlations of the observables for future work. 

To finish, we provided an expression for the direction and redshift drifts in Bianchi I universes in the ``straight geodesic approximation". A multipolar decomposition shows that the shear contributes as a quadrupole, which makes it distinguishable from the effects of the velocity. We also showed that the results published in the literature were actually missing an aberration drift type contribution. We then estimated how well the shear can be constrained using our results.

\section*{Acknowledgements} OM thanks IAP for warm hospitality during this work. We thank Christian Marinoni for discussions and comments on the draft, Pierre Fleury for his insight concerning the relation~(\ref{tzrelation}) and Guilhem Lavaux for discussions on the local universe. The work of OM was supported by the PDSE program by CAPES. The work of CP and JPU is made in the ILP LABEX (under reference ANR-10-LABX-63) and was supported by French state funds managed by the ANR within the Investissements d'Avenir programme under reference ANR-11-IDEX-0004-02.  JPU acknowledges the financial support from the EMERGENCE 2016 project, Sorbonne Universit\'es, convention no. SU-16- R-EMR-61 (MODOG). TSP is supported by the Brazilian research agencies CNPq and Fundação Araucária.

\bibliography{driftdraft}

\appendix

\section{Time-redshift relationship} \label{apendixA}
Here we show the validity of the relation
\begin{equation}
\frac{\dd \tau_{s}}{\dd \tau_{\obs}}=\frac{\omega_{\obs}}{\omega_{s}}=\frac{1}{1+z}
\end{equation}
where $\tau_{s,\obs}$ are the proper times of the source and the observer, respectively. In the geometrical optics approximation (also known as eikonal approximation), the electromagnetic vector potential satisfies (in vacuum)
\[\label{app1} \nabla^{\mu}\nabla_{\mu}A_{\nu}=0. \]
Considering a solution of the form
\[\label{app2} A_{\mu}=C_{\mu}e^{i\varphi}\]
with approximately constant amplitude, Eq.~\eqref{app2}, neglecting derivatives of $C_{\mu}$, gives
\begin{align}
\nabla^{\mu}\varphi \nabla_{\mu}\varphi&=0 \label{eA4} \\
\nabla^{\mu} \nabla_{\mu} \varphi&=0.
\end{align}
$k^{\mu}=\nabla^{\mu}\varphi$ is the vector normal to the surfaces of constant $\varphi$. Differentiating (\ref{eA4})  gives the geodesic equation
\[k^{\mu}\nabla_{\mu}k_{\nu}=0.\]
Hence $k^{\mu}$ is also tangent to a null geodesic. Thus, in the eikonal approximation, null geodesics are curves of constant phase $\varphi$. 

The frequency of the wave as measured by an observer with 4-velocity $u^{\mu}$ is precisely (minus) the rate of change of the phase of the wave with respect to his proper time. That is,
\[\omega=-\frac{\dd\varphi}{\dd\tau}=-u^{\mu}\nabla_{\mu}\varphi=-u^{\mu}k_{\mu}.\] Thus, for a source and observer connected by a null geodesic, we have that
\[\frac{\omega_{s}}{\omega_{\obs}}=\frac{\dd\varphi/\dd\tau_{s}}{\dd\varphi/\dd\tau_{\obs}}.\]
Since null geodesics are curves of constant phase, we thus have that
\begin{equation} \label{tzrelation}
\frac{\dd\tau_{s}}{\dd\tau_{\obs}}=\frac{\omega_{\obs}}{\omega_{s}}=\frac{1}{1+z}.
\end{equation}
Note that in the cases of inertial FL and Bianchi I the observer and source were both comoving so that their proper times coincide with the cosmic time. For a general observer, this holds only to first order in velocities. For a perturbed FL, the difference between proper time and cosmic time has to be taken into account.

\section{Drift of integrated effects in perturbed FL} \label{apendixB}

Here we show how to evaluate $\delta_{12}\Xi$ where $\Xi$ is a general term of the form $\Xi=\int_{\eta_\obs}^{\eta_s}\xi[\eta,x^i]\dd\eta$. 

To that purpose, one has to take into account how $\xi[\eta,x^i]$ changes from one geodesic to the other. Thus, there will be a contribution due to the fixed endpoints of the integral but also a contribution from the spatial change of the integrand. More explicitly,
\begin{equation}
\delta_{12}\Xi=\int_{\eta_\obs+\delta\eta_\obs}^{\eta_s+\delta\eta_s}\xi[\eta,x^i_2]\dd\eta-\int_{\eta_\obs}^{\eta_s}\xi[\eta,x^i_1]\dd\eta.
\end{equation}
The second line of sight $x^i_2(\eta)$ is given by
\begin{align}
x^i_2(\eta)&=x^i_1(\eta)+\delta x^i(\eta) \nonumber\\
&= x^i_1(\eta)-\bar{n}^i\delta\eta.  \label{appB1}
\end{align}
This allows us to compute the contribution from the endpoints, which, to first order in $\delta\eta$, is simply
\begin{align}
\int_{\eta_\obs+\delta\eta_\obs}^{\eta_s+\delta\eta_s}\xi[\eta,x^i_2]\dd\eta=&\int_{\eta_\obs}^{\eta_s}\xi[\eta,x^i_2]\dd\eta\\
&+\delta\eta_s\xi_s[x^i_1] -\delta\eta_\obs\xi_\obs[x^i_1] .\nonumber
\end{align}
Now, we use Eq.~\eqref{appB1} to write 
\begin{equation}
\xi[\eta,x^i_2]=\xi[\eta,x^i_1]-\bar{n}^i\partial_i \xi[\eta,x^i_1]\delta\eta\,.
\end{equation}
Now, remember that $\bar{n}^i\partial_i=\frac{\dd}{\dd\eta}-\frac{\partial}{\partial\eta}$. Also, since $\xi$ is a perturbation, any term multiplying it is evaluated at the background, and at this level $\delta\eta_s=\delta\eta_\obs$. We conclude that
\begin{equation}
\delta_{12}\Xi=\delta\eta_\obs\,\int_{\eta_\obs}^{\eta_s}\xi^\prime[\eta,x^i_1]\dd\eta\,.
\end{equation}

\end{document}